\documentclass{PoS}

\usepackage{wrapfig}

\title{Neutrino-nucleus DIS data and their consistency with nuclear PDFs}

\ShortTitle{Neutrino-nucleus DIS data and their consistency with nuclear PDFs}

\author{\speaker{Hannu Paukkunen}%
          \\
          University of Jyv\"askyl\"a, Department of Physics, Finland \\ 
          Helsinki Institute of Physics, Finland \\
        E-mail: \email{hannu.paukkunen@jyu.fi}}

\author{Carlos A. Salgado\\
        University of Santiago de Compostela,
        Departamento de F\'isica de Part\'iculas and IGFAE, Spain \\
        E-mail: \email{carlos.salgado@usc.es}}

\abstract{In this talk, we discuss the compatibility of different deeply inelastic neutrino-nucleus
data sets and the universal nuclear PDFs. This is an issue that has
lately been investigated by different groups but the conclusions have been surprisingly contradictory. While some
studies have found a good overall agreement between the nuclear PDFs and the neutrino data, others have
claimed for an incompatibility. Here, we demonstrate that the independent neutrino data sets from NuTeV,
CHORUS and CDHSW collaborations differ in the absolute overall normalization and that it is not possible
to accurately reproduce all the data simultaneously with a single set of PDFs. Our strategy to overcome
this difficulty and allow a consistent use of all neutrino data in global PDF analyses is to normalize
the data by the integrated cross-sections thereby cancelling possible inaccuracies in the absolute normalization.
Indeed, this brings all data to a surprisingly good mutual agreement underscoring the x-dependence of the
nuclear modifications in a model-independent way. The consistency of these data with the present nuclear PDFs
is verified by introducing a method to test the effect of a new data set in an existing global fit that performed
a Hessian error analysis.}

\FullConference{XXI International Workshop on Deep-Inelastic Scattering and Related Subjects\\
                 22-26 April, 2013\\
                 Marseilles, France}

\begin{document}

\section{Introduction}

The large-$Q^2$ neutrino-nucleus ($\nu A$) deep inelastic scattering is an interesting ``cross-over''
process that can shed light on quite different sides of parton distribution functions (PDFs).
In leading order, the partonic content of the nucleon probed in charged-current $\nu A$ cross-sections can be schematically 
written as
$$
\frac{d^2\sigma^{\nu A}}{dxdy}  \propto \left( d + s + b \right) + \left(1-y \right)^2 \left( \overline{u} + \overline{c} \right)
\quad ; \quad
\frac{d^2\sigma^{\overline{\nu} A}}{dxdy} \propto \left( \overline{d} + \overline{s} + \overline{b} \right) +
 \left(1-y \right)^2 \left( u + c \right), 
$$
which should be compared to the corresponding expression for the standard charged-lepton induced neutral-current process
$$
\frac{d^2\sigma^{\ell^\pm A}}{dxdy} \propto \frac{4}{9} \left( u + c + \overline{u} + \overline{c} \right) +
\frac{1}{9} \left( d + s + b + \overline{d} + \overline{s} + \overline{b} \right).
$$
Of importance to the free proton analyses is the strange quark PDF that is more
pronounced in neutrino reactions than in the charged-lepton process where the already 
small strange quark PDF is additionally suppressed by the QED coupling. On the other
hand, the neutrino data is taken with nuclear targets and should therefore be useful 
for constraining the nuclear effects in PDFs.

The use of $\nu A$ data for either purpose relies naturally on the adequacy
of the collinear factorization in these processes which many free proton fits
take as granted by using these data. However, this assumption has been doubted. In particular, it was
reported \cite{Schienbein:2007fs,Schienbein:2009kk} that the $\nu A$ data from the NuTeV collaboration
\cite{Tzanov:2005kr} prefer quite different nuclear modifications in PDFs as the other existing $\ell^\pm A$ data.
Later publication \cite{Kovarik:2010uv} by the same collaboration declared all $\nu A$ data as incompatible
with the $\ell^\pm A$ data. 
Rather different strategy was adopted in \cite{Paukkunen:2010hb}, where data from independent neutrino experiments
(NuTeV \cite{Tzanov:2005kr}, CDHSW \cite{Berge:1989hr}, CHORUS \cite{Onengut:2005kv}) were contrasted with the existing
nuclear PDFs. While an excellent overall global agreement was found, surprisingly large, beam energy dependent 
fluctuations in the absolute normalization of the NuTeV data sample were noticed and suggested to cause the results of
\cite{Schienbein:2007fs,Schienbein:2009kk,Kovarik:2010uv}. In a recent analysis \cite{deFlorian:2011fp},
these $\nu A$ data were included in a global fit of nuclear PDFs. No difficulties in accommodating these data with other
$\ell^\pm A$ measurements was reported. However, this analysis differs from the others
in utilizing the structure functions extracted by the experiments instead of the absolute cross-sections.
Also, the uncertainties from the baseline PDFs were added on top of the experimental errors. 
Here, we review the results of the latest effort \cite{Paukkunen:2013grz} --- inspired by the findings of \cite{Paukkunen:2010hb} ---
that sidesteps the possible experimental issues in the absolute normalization. 

\vspace{-0.2cm}
\section{Experimental Input And The Theoretical Framework}

\vspace{-0.2cm}
The experimental neutrino cross-sections that enter to the analysis come from three independent
(Fermilab and CERN) experiments: NuTeV \cite{Tzanov:2005kr}, CDHSW \cite{Berge:1989hr} and CHORUS \cite{Onengut:2005kv}.
After applying typical cuts for the virtuality $Q^2 > 4 \, {\rm GeV}^2$ and for
the invariant mass of the final state $W^2 > 12.25 \, {\rm GeV}^2$, 2136 NuTeV, 824 CHORUS and 937 CDHSW 
data points remain. The neutrino beam energy ranges from $E \sim 20 \, {\rm GeV}$ up to $E \sim 300 \, {\rm GeV}$.
As in \cite{Schienbein:2007fs,Schienbein:2009kk,Kovarik:2010uv,Paukkunen:2010hb}, the theoretical calculations
are performed at next-to-leading order pQCD supplemented with the SACOT prescription for the treatment of
heavy quarks (in \cite{deFlorian:2011fp} a different scheme was adopted). Accordingly, we utilize the
CTEQ6.6 \cite{Nadolsky:2008zw} free proton PDFs, and the EPS09 nuclear modifications \cite{Eskola:2009uj}.
Corrections for electroweak radiation and target-mass effects are applied \cite{Paukkunen:2010hb}.

\section{The Normalization Procedure}

\begin{figure*}[ht]
\begin{minipage}[b]{0.45\linewidth}
\centering
\includegraphics[width=\textwidth]{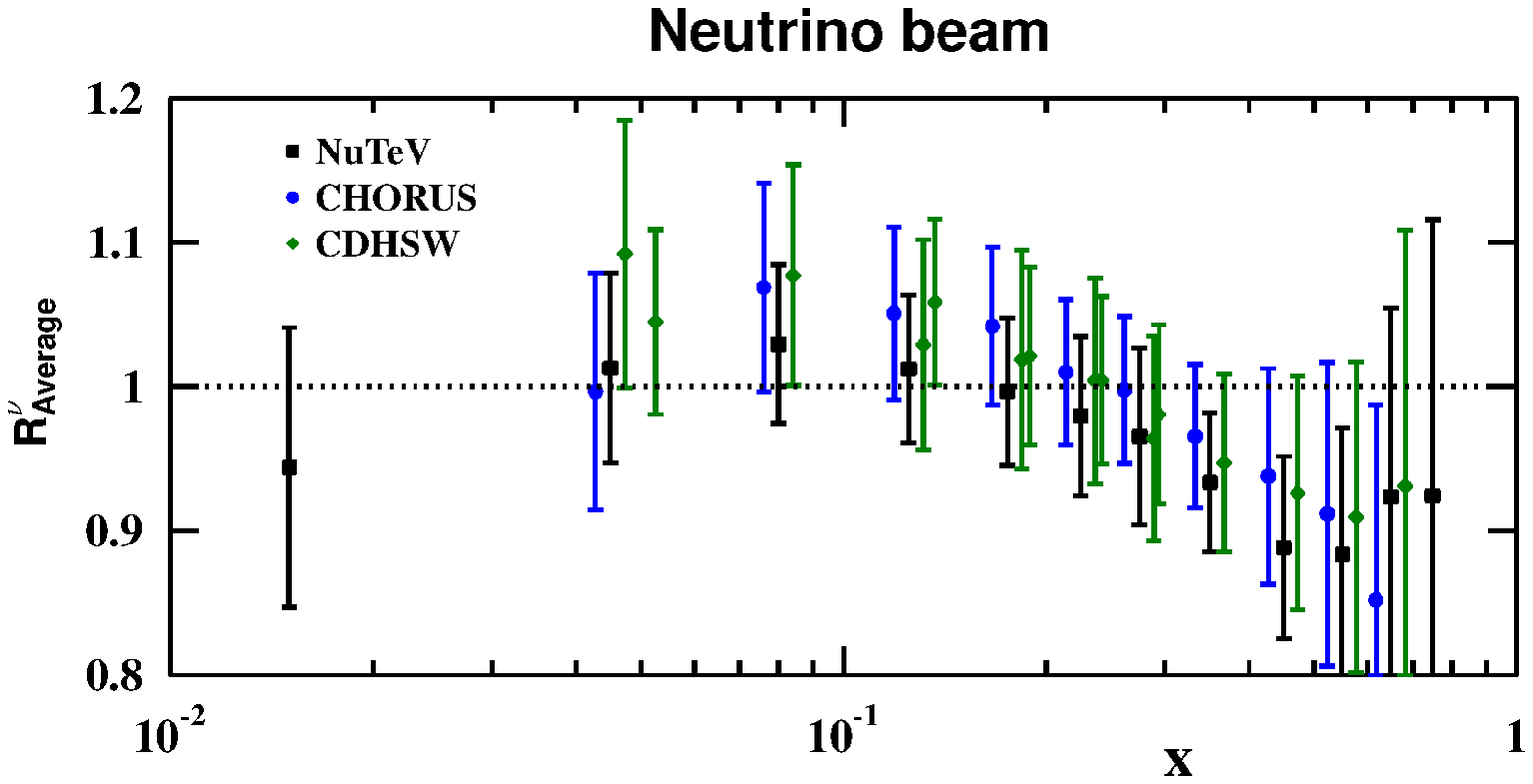}
\end{minipage}
\hspace{0.5cm}
\begin{minipage}[b]{0.45\linewidth}
\centering
\includegraphics[width=\textwidth]{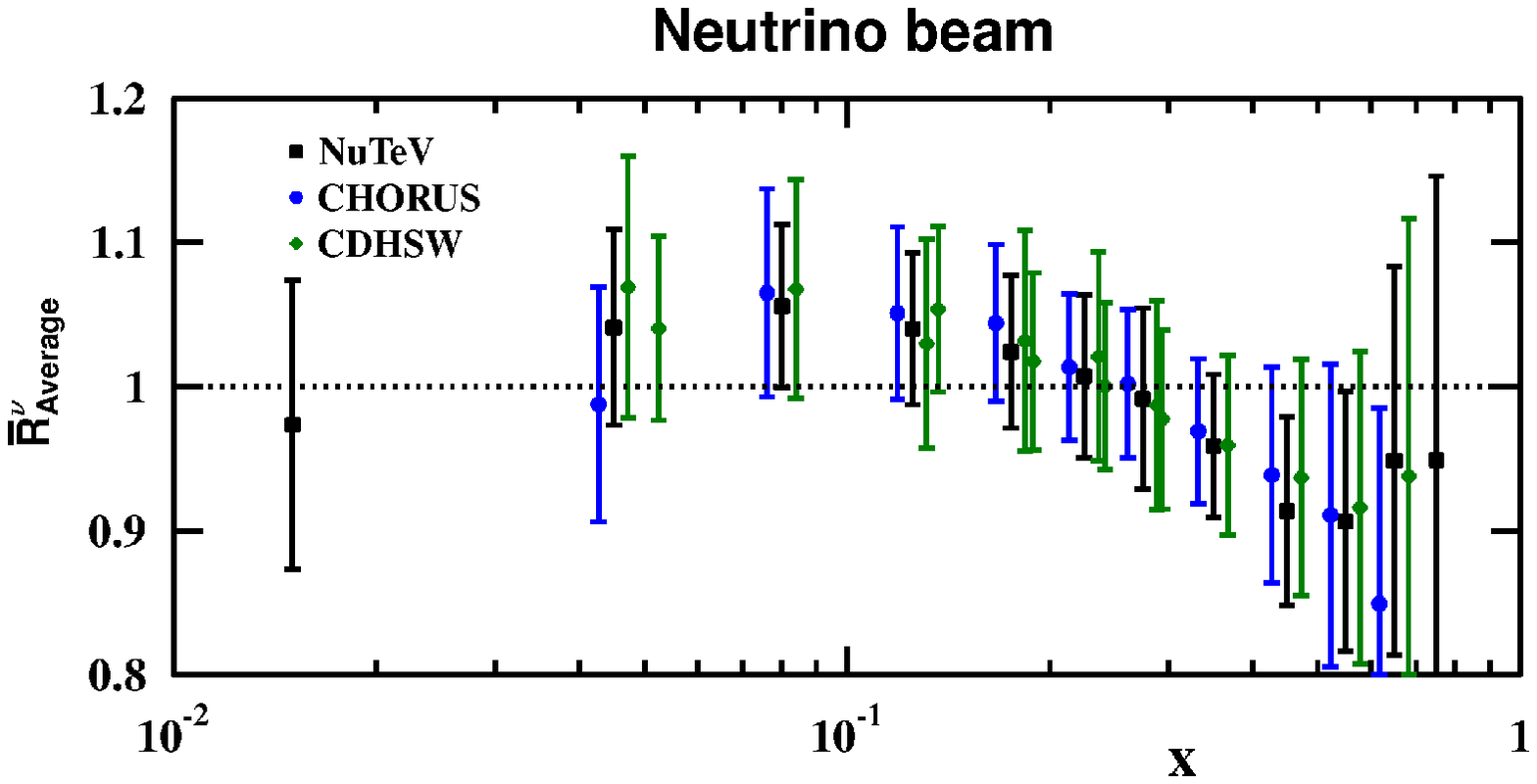}
\end{minipage}
\caption{The neutrino data presented as $R^{\nu}_{\rm Average}$ (left panel),
and as $ \overline R^{\nu}_{\rm Average}$ (right panel). The CHORUS (blue circles) and CDHSW (green diamonds)
data has been horizontally shifted from the NuTeV (black squares) data points.}
\label{Fig:Neutrino1}
\end{figure*}

\begin{wrapfigure}{r}{0.45\textwidth}
\centerline{\includegraphics[width=0.45\textwidth]{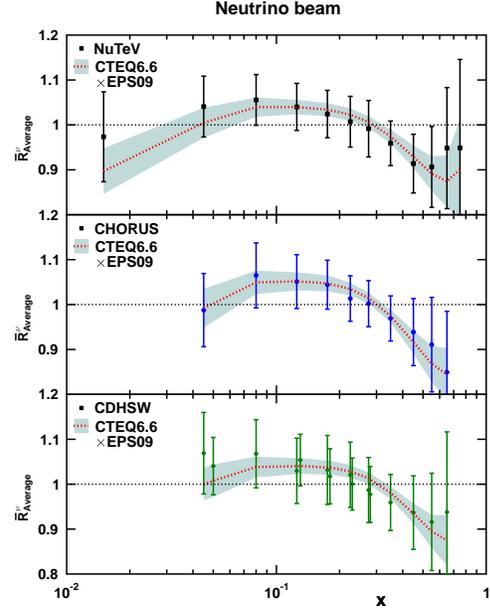}}
\caption{The experimental $\overline R^{\nu}_{\rm Average}$ compared to the predictions from CTEQ6.6$\, \otimes\, $EPS09.}
\label{Fig:PbPDFs}
\end{wrapfigure}

Instead of comparing the calculations directly with the absolute experimental cross-sections $\sigma^{\nu}_{\rm exp}(x,y,E)$,
we form a ratio
\begin{equation}
 R^{\rm \nu}(x,y,E) \equiv \frac{\sigma^{\nu}_{\rm exp}(x,y,E) }{\sigma^{\nu}_{\rm CTEQ6.6}(x,y,E)}, \label{eq:StandardR}
\end{equation}
where $\sigma^{\nu}_{\rm CTEQ6.6}(x,y,E)$ is calculated without nuclear effects in PDFs.
This facilitates the interpretation of the data vs. theory comparison. As
found in \cite{Paukkunen:2010hb}, these ratios are practically independent of $Q^2$
and the beam energy $E$. Therefore, we construct a following weighted average
\begin{equation}
 R^{\nu}_{\rm Average}(x) \equiv
 \left( \sum^N_{i\in {\rm fixed} \, x} \frac{R_i^{\nu}}{\delta_i} \right) \left( \sum^N_{i\in {\rm fixed} \, x} \frac{1}{\delta_i} \right)^{-1}
 \pm
 N \times \left( \sum^N_{i\in {\rm fixed} \, x} \frac{1}{\delta_i} \right)^{-1},
\end{equation}
where $\delta_i$ stands for the experimental error (divided by $\sigma^{\nu}_{\rm CTEQ6.6}$) and $N$ is the number of data points.
This procedure effectively 
distills the average value of $R^{\rm \nu}(x,y,E)$ for a given $x$-bin and gives an idea of its uncertainty.
The left-hand panel of
Figure~\ref{Fig:Neutrino1} presents the results obtained in this way. Non-negligible
differences in the absolute normalization are visible. Especially, the NuTeV neutrino
data is systematically below the rest.
Apart from these normalization differences the $x$ dependence of the $ R^{\nu}_{\rm Average}$
appears quite similar for each experiment. Motivated by this observation, we define
\begin{equation}
 I_{\rm exp}^\nu(E) \equiv \sum_{i\in {\rm fixed} \, E} \sigma_{{\rm exp},i}(x,y,E) \times B_i(x,y),
\end{equation}
where $B_i(x,y)$ is the size of the experimental $(x,y)$-bin. That is,  $I_{\rm exp}^\nu(E)$
is an estimate for the integrated cross-section in an energy bin. Now, instead of Eq.~(\ref{eq:StandardR})
we consider
\begin{equation}
 \overline R^{\rm \nu}(x,y,E) \equiv \frac{\sigma^{\nu}_{\rm exp}(x,y,E)/I^\nu_{\rm exp}(E)}{\sigma^{\nu}_{\rm CTEQ6.6}(x,y,E)/I^\nu_{\rm CTEQ6.6}(E)}.
  \label{eq:SNewR}
\end{equation}
The right-hand panel of Figure~\ref{Fig:Neutrino1} show how this simple normalization 
procedure seems to bring all data in perfect mutual agreement. In Figure~\ref{Fig:PbPDFs},
we show a comparison with the theoretical predictions from the nuclear PDFs
defined in the usual manner as
\begin{equation}
f_{i}^A(x,Q^2) \equiv R_{i}^{A, {\rm EPS09}}(x,Q^2) f_{i}^{\rm CTEQ6.6M}(x,Q^2).  \label{eq:nPDF}
\end{equation}
The data points are the same as in Figure~\ref{Fig:Neutrino1}, and the blue band represents the uncertainty range
derived using the CTEQ6.6 and EPS09 error sets. We note that large part of the CTEQ6.6 uncertainty
cancels in the normalization procedure. Clearly, the nuclear PDFs can reproduce the normalized data.
For the corresponding figures in the case of the antineutrino data, see \cite{Paukkunen:2013grz}.

\vspace{-0.3cm}
\section{The Numerical Check}

\vspace{-0.3cm}
We verify the consistency of these data within the CTEQ6.6 and EPS09 global fits by
the Hessian reweighting technique \cite{Paukkunen:2013grz} \footnote{An article on
the relation and differences to the NNPDF reweighting \cite{Ball:2010gb} and MSTW work \cite{Watt:2012tq}
will appear later.}.
The method relies on the PDF uncertainty sets $S_k^\pm$ that quantify the $\Delta \chi^2$ neighborhood of the best fit $S_0$
found in a global analysis. These sets can be used to estimate the values of any PDF-dependent
quantities $X_k$ close to the best fit as
$$
 X_k \left[S \right] \approx X \left[S_0 \right] + \sum_k \frac{\partial X_k \left[S \right]}{\partial z_k}{\Big|_{S=S_0}} z_k
                   \approx X_k \left[S_0 \right] + {\bf D}_k \cdot {\bf w}, \label{eq:XS}
$$
where $(D_k)_i \equiv ({X_k\left[S_i^+ \right] - X_k\left[S_i^- \right]})/{2}$ and $w_i \equiv {z_i}/ \sqrt{\Delta \chi^2}$.
Here, we take the $X_k$ as the neutrino cross-sections and study their
compatibility within the global fits by defining a $\chi^2$ function by
\begin{equation}
 \chi^2 \equiv \sum_{\{X^{\rm data}\}} \left[ \frac{X_k\left[S\right] - X_k^{\rm data}}{\delta_k^{\rm data}} \right]^2 + 
 \Delta \chi^2_{\rm EPS09} \sum_{k=1}^{15} w_k^2 + \Delta \chi^2_{\rm CTEQ6.6} \sum_{k=16}^{37} w_k^2, \label{eq:newchi2}
 \end{equation}
where $\Delta \chi^2_{\rm EPS09}=50$ and $\Delta \chi^2_{\rm CTEQ6.6}=100$. This expression is
a quadratic function of the parameters $w_i$ and its minimum can be found by the standard methods 
of linear algebra. The values of the ``penalty terms`` $\Delta \chi^2_{\rm EPS09 \backslash CTEQ6.6} \sum_k w_k^2$ at
the minimum can be used to distinguish whether the new data set is in agreement with the original fits:
If $\Delta \chi^2_{\rm EPS09 \backslash CTEQ6.6} \sum_k w_k^2 \ll \Delta \chi^2_{\rm EPS09 \backslash CTEQ6.6}$,
the new data agrees well with the original fit but if $\Delta \chi^2_{\rm EPS09 \backslash CTEQ6.6} \sum_k w_k^2 \gtrsim \Delta \chi^2_{\rm EPS09 \backslash CTEQ6.6}$
tension clearly exists.

The Table~\ref{Table:Data} displays the key results in the case of the NuTeV data. The first column $\chi^2_{w=0}/N$
corresponds to the $\chi^2$ calculated by the central values from CTEQ6.6$\, \otimes\, $EPS09 (zero penalty). 
The normalization clearly improves the agreement. The next column $\chi_{w_{\rm min}}^2/N$ shows
what happens when the minimization is performed. This naturally improves
the agreement. However, this also gives rise to the penalty terms and if no normalization is applied,
the penalty for the EPS09 is already close to the largest permitted value 50. With the normalization, the
penalties remain small which indicates that the normalized NuTeV data could be added to these global fits.
For the CHORUS and CDHSW data the penalties remain 
always very small (see \cite{Paukkunen:2013grz}). 
\begin{table*}
\begin{center}
{\footnotesize
\begin{tabular}{rcccc||cc}
& \multicolumn{4}{c||}{All CTEQ6.6 and EPS09 error sets} & \multicolumn{2}{c}{Only EPS09 error sets} \\
&       &           &           &       &            &     \\  
NuTeV \vline & $\chi^2_{w=0}/N$ & $\chi_{w_{\rm min}}^2/N$ & EPS09-penalty & CTEQ-penalty &  $\chi_{w_{\rm min}}^2/N$ & EPS09-penalty \\
Normalization \vline &  0.84     &    0.77        &     13.9      &    35.4      &    0.81        &     33.8 \\
No normalization \vline &  1.04     &    0.90        &     40.3      &    42.5   &    0.94        &     77.4 \\
\end{tabular}
}
\caption[]{\small The $\chi^2/N$ for the NuTeV data and the EPS09 and CTEQ6.6 penalties. See the text for details.
}
\label{Table:Data}
\end{center}
\end{table*}
In order to mimic the analysis of Ref.~\cite{Kovarik:2010uv} where an incompatibility was found,
we freeze the CTEQ6.6 to its central value. The corresponding results (now the CTEQ penalty is zero) values are shown in the
two right-most panels. Without the normalization, the EPS09-penalty is almost 80 --- clearly
above the permitted 50. That is, we would reach the same conclusion as the authors of Ref.~\cite{Kovarik:2010uv}.

\vspace{-0.3cm}
\section{Conclusion}

\vspace{-0.3cm}
As a summary, we have demonstrated that independent neutrino data sets seem to disagree
in the absolute normalization. Especially, the NuTeV data show a difference with
the rest. We propose to normalize the data by the corresponding integrated
cross-section which appears to largely dispose the differences among the data sets.
The Hessian reweighting technique is used to study the consistency
with the present nuclear PDFs, and a good agreement is found when the normalization
procedure is considered. Without the normalization we recover the contradictory
results of Ref.~\cite{Kovarik:2010uv}.

\end{document}